\documentclass [12pt]{article}
\usepackage{fullpage}
\usepackage{graphicx} 
\usepackage{setspace}
\begin{document}

\title{A Problem-Specific Fault-Tolerance Mechanism for Asynchronous,
Distributed Systems}

\author{Adriana Iamnitchi\thanks{Department of Computer Science, The
University of Chicago, Chicago, IL 60637} \and Ian
Foster\thanks{Mathematics and Computer Science Division, Argonne
National Laboratory, Argonne, IL 60439}\ \footnotemark[1]}

\date{}
\maketitle

\begin{abstract}

The idle computers on a local area, campus area, or even wide area
network represent a significant computational resource---one that is, however,
also unreliable, heterogeneous, and opportunistic. This type of
resource has been used effectively for embarrassingly parallel
problems but not for more tightly coupled problems. We describe an
algorithm that allows branch-and-bound problems to be solved in such
environments. In designing this algorithm, we faced two challenges:
(1) scalability,  to effectively exploit the variably sized pools of
resources available, and (2) fault tolerance, to ensure the
reliability of services. We achieve scalability through a fully
decentralized algorithm, by using a membership protocol for managing
dynamically available resources. However, this fully decentralized
design makes achieving reliability even more challenging. We guarantee fault
tolerance in the sense that the loss of 
up to all but one resource will not affect the quality of the
solution. For propagating information efficiently, we use epidemic
communication for both the membership protocol and the fault-tolerance
mechanism. We have developed a simulation framework that allows us 
to evaluate design alternatives. Results obtained in this framework
suggest that our techniques can execute scalably and reliably.
\end{abstract}

\section{Introduction}

For solving new, more difficult search problems, scientists
need better search heuristics and/or more powerful resources. The need
for hundreds or even thousands of processors is justified in the case of
branch-and-bound search algorithms by problems that could not be
solved after months of execution on tens of processors \cite{eckstein94,
hahn}.

Rarely, however, are thousands of processors assembled in a single
location and available for a single problem.
Thus, techniques are needed that would allow us to aggregate processors
at many different Internet-connected locations.  These processors are
likely often to be required for other purposes; hence their availability
will be episodic, and any algorithm designed to take advantage of these
resources must be opportunistic.  Furthermore, the Internet environment
is likely to be unreliable and heterogeneous.

Various groups have demonstrated the feasibility of using Internet-connected
computers for solving embarrassingly parallel problems \cite{lenstra,
condor}. In our work, we investigate the 
feasibility of 
applying Internet-connected resources to more tightly coupled problems, in
which a centralized scheme is not computationally efficient.
Our approach is to develop specialized algorithms that incorporate 
scalability and reliability mechanisms.

For providing reliable services over unreliable architectures,
researchers usually choose one of the following approaches: (1) embed
 fault-tolerance mechanisms within the middleware software layer, as
in ISIS \cite{isis} or CORBA Transaction Service, or as in systems
like Condor \cite {chkpt-mngt, hjk} or Legion \cite{legion}; or (2)  
embed fault-tolerance mechanisms within algorithms. The former
approach is more general. Successful 
results  in this domain guarantee communication and hardware
reliability to a large number of 
applications. But its generality imposes problems that
sometimes turn out to be unsolvable \cite{fischer, chandra} or very
expensive. The latter alternative is applicable to specific problem
classes and is therefore less general. But exploiting the characteristics
of a class of problems may ease the design of fault-tolerance
mechanisms, yielding simpler and more efficient algorithms. Note that
middleware can still be of assistance in this case, by providing
appropriate fault-detection services \cite{ft-globus}.

In our work, we focus on problem-specific fault-tolerance
mechanism. Specifically, we propose a fault-tolerant, totally
distributed branch-and-bound 
algorithm designed for unreliable architectures, with a dynamically
variable number of resources. The description of the
branch-and-bound problem (Section \ref{sec:BB}) and the target
architecture (Section  
\ref{sec:Architecture}) provide the motivation for our work. We
describe our branch-and-bound algorithm in Section
\ref{sec:Algorithm}, focusing particularly on the
fault-tolerance mechanism. Related work (Section \ref{sec:Related}) refers
 to other fault-tolerance techniques embedded in tree-based,
distributed, asynchronous algorithms. For testing our solution, we
have developed a simulation framework, which is presented in Section
\ref{sec:Simulation}, along with the results obtained. We
conclude with a discussion of what we learned from trying to solve
this problem and how we intend to continue this work.

\section{Branch and Bound}
\label{sec:BB}
The search for optimal solutions is one of the most important
searching problems. Since exhaustive search is often 
impracticable in NP-hard problems, heuristics are employed to improve search
performance. Branch-and-bound (which we will hereafter
refer to as B\&B) is an intelligent search method often used for optimization problems. It uses a successive decomposition
of the original problem into smaller disjoint subproblems, while
reducing (\textit{pruning}) the search space by recognizing
unpromising problems before starting to solve them.

A sequential B\&B algorithm consists of a sequence of iterations in
which four basic operators are applied over a list of problems, called a 
\textit{pool of active problems}:
\begin{itemize}
\item [a.]
\textbf{Decompose}. Splits a problem into a set of new
subproblems. A problem that cannot be split (either because it
has no solution or because a solution is found) is \textit{fathomed}. A
problem decomposed into new subproblems is \textit{branched}. 
\item [b.]
\textbf{Bound}. Computes a bound value $l(v)$ on the optimal solution of
subproblem $v$. This bound value will be used by \textit{Select} and
\textit{Eliminate} operations.
\item [c.]
\textbf{Select}. Selects which problem to branch from next, as a function
of some
heuristic priority function. Selection may depend on bound values,
such as in the \textit{best-first} selection rule, or not, as in the case
of \textit{depth-first} or \textit{breadth-first} rules.
\item [d.]
\textbf{Eliminate}. Eliminates problems that cannot lead to an
optimal solution of the original problem (i.e.,  problems for which
$l(v) \ge U $, where $U$ is the \textit{best known solution}).
\end{itemize}

Successive decomposition operations create a tree of problems rooted
in the original problem. The value of the best solution found thus far
is used to 
recognize the unpromising problems and \textit{prune} the tree. If the
bound value of the current problem is not better than the best-known
solution, then the problem is eliminated. Otherwise, it is stored into
the pool of active 
problems. The best-known solution is updated when a better feasible
solution is found. The leaves of the tree are infeasible 
problems, or pruned problems, or problems that lead to locally optimal
solutions. The size and shape of the tree strongly depends on the quality of
the heuristic function for the selection rule. 

In B\&B algorithms, parallelism can be achieved in different ways
\cite{crainic94}. We consider the most general approach, in which the B\&B
tree is built in parallel by  performing operations on different
subproblems simultaneously. 

Three design choices most influence the performance of
parallel B\&B algorithms: the choice of a synchronous or an
asynchronous algorithm, the \textit{work sharing} mechanism, 
and the \textit{information sharing} mechanism. Synchronous
vs. asynchronous design defines
what processes do upon completion of a work unit---they wait for each
other (in the case of synchronous algorithms) or not (in asynchronous
algorithms). Work sharing is the
method used to assign work to processes in order to fully and
efficiently exploit available parallelism. Information sharing refers to the
methods used to publish and update the best-known solution. Using an
up-to-date best-known solution improves the efficiency of the selection
and elimination rule  and hence has an important effect on the size of
the search space.

\section{Related Work}
\label{sec:Related}

Many investigations of parallel B\&B for distributed-memory systems have
adopted a centralized 
approach in which a single manager maintains the tree and
hands out tasks to workers \cite{crainic94, BBtaxonomy}. While
clearly not scalable, this approach simplifies the management of
information and multiple processes. Scalability can be
improved through a hierarchical organization of processes or by
varying the 
size of work units, but the central manager remains an obstacle to both
scalability and fault tolerance. Reliability can be achieved
through checkpointing, but this approach assumes that there exists at
least one reliable process/machine, able to manage the failure
recovery process. 

Because of the highly variable number of resources in the architecture
we consider, we need more flexibility than that offered by the
centralized design. Hence we chose a fully decentralized design. 

The only fully decentralized, fault-tolerant B\&B algorithm for
distributed-memory architectures is DIB (Distributed Implementation of Backtracking)
\cite{dib}. DIB was  designed for a wide range of tree-based
applications, such as recursive backtrack, branch-and-bound, and
alpha-beta pruning. It is a distributed, asynchronous algorithm
that uses a dynamic load-balancing technique. Its failure recovery
mechanism is based on keeping track of which machine is responsible
for each unsolved problem. Each machine memorizes
the problems for which it is responsible, as well as the machines to
which it sent problems or from which it received problems. The completion
of a problem is reported to the machine the problem came from. Hence,
each machine can determine 
whether the work for which it is responsible is still unsolved, and
can redo that work in the case of failure. 

\section{Target Architecture}
\label{sec:Architecture}
The target architecture for our algorithm is a collection of Internet-connected
computers. The distinctive characteristics of this environment,
when compared with a conventional parallel computer, are as
follows:  

\begin{itemize}

\item [] \textbf{Scale}. The number of resources available can potentially
be much larger than on a conventional parallel computer.

\item [] \textbf{Dynamic availability}. The quantity of resources available
may vary over time, as may the amount of computation delivered by a
single resource. 

\item [] \textbf{Unreliability}. Resources may become unreachable
without notice because of system or network failures. 

\item [] \textbf{Communication characteristics}. Latencies may be high,
variable, and unpredictable; bandwidth may be low, variable, and
unpredictable. Connectivity (as measured, for example, by bisection
bandwidth) may be particularly low.

\item [] \textbf{Heterogeneity}. Resources may have varying physical
characteristics (for example, amount of memory, speed). 

\item [] \textbf{Lack of centralized control}. There is no central
authority for quality control or operational management.
\end{itemize}

The failure model we consider is Crash \cite{cristian, schneider}, in which
a processor fails by halting. Once it halts, the processor remains in
that state. The fact that a processor has failed may not be detectable
by other processors. We make minimum assumptions about the system: 

 -- There is no bound on message delivery time. 

 -- Messages may be lost altogether. 

 -- A network link does not duplicate, corrupt, or spontaneously create
messages.

 -- The clock rate on each host is close to accurate (we do not assume
that the clocks are synchronized). This condition is assumed in many
works in the fault-tolerance domain and it does not represent a
\textit{practical} restriction \cite{cris-fetzer}.

\section{The Algorithm}
\label{sec:Algorithm}
We propose a fully decentralized, asynchronous, fault-tolerant
parallel B\&B algorithm suited for the environment described
above. Asynchrony is required by the heterogeneity of the architecture  
and allowed by the B\&B problem \cite{crainic94}. Each process maintains its
local pool of problems to be solved. When the local pool is empty, the
process sends work requests to other processes. A process that
receives a work request and has enough problems in its pool removes
some of those problems and sends them to the requester. This on-demand
dynamic load-balancing scheme was chosen to reduce unnecessary
communication. 
The fully decentralized scheme was preferred for better scalability
and for greater reliability. The information sharing issue is solved
by circulating the best-known solution among processes, embedded in
the most frequently sent messages. Processes update the local value
for the best-known solution every time they receive it, and use it
when the next decision is to be made. 

For adapting this rather conventional B\&B algorithm to the
environment described above, we 
extend it with (1) a group membership protocol to allow
dynamic variation in the number of resources and (2) a fault-tolerance
mechanism. The novelty of this paper is the
decentralized fault-tolerance mechanism that uses a tree-based
encoding of the B\&B subproblems. This strategy for problem encoding
also offers  a simple mechanism for termination detection, described in
Section \ref{sec:TD}. A brief description of the epidemic communication mechanism
(Section \ref{sec:EC}) will help in understanding how the group membership
protocol (Section \ref{sec:GMP}) and fault-tolerance mechanism
(Section \ref{sec:FTM})
function. A comparison with DIB, the decentralized B\&B algorithm
mentioned in Section \ref{sec:Related}, concludes this section.

\subsection{Epidemic Communication for Group Membership and Fault
Tolerance}
\label{sec:EC}

Epidemic communication \cite{alon}
allows temporary inconsistencies in shared data in exchange for
low-overhead implementation. More specifically, information changes are
spread gradually throughout the processes, without the overhead and
communication costs typically used to achieve a high degree of
consistency. 

Both our group
membership and fault-tolerance mechanisms use epidemic
communication. Since these mechanisms do not require data consistency,
epidemic communication is a convenient algorithm for spreading 
information. However, epidemic communication guarantees that
eventually consistency is achieved; that is, all processes will
eventually see the same data when no more new information is brought
into the system, independent of system failures \cite{farley80,
laforest97}. This observation is exploited for termination detection. 

The epidemic algorithms used are variants of
the \textit{rumor-mongering} algorithm (analyzed in \cite{demers88}):
when a site receives a new update  
(\textit{rumor}), it becomes ``infectious'' and is willing to share---it
repeatedly chooses another member, to which it sends the rumor. Upon
receipt of a rumor, a member updates its local information and
sends its own version after some time interval. In the membership
protocol, the rumor received is sent farther, without being
processed. In the fault-tolerance mechanism, the rumor is stored for
local processing, may be processed locally, and is spread
infrequently. 

\subsection{Group Membership Protocol}
\label{sec:GMP}

The group membership protocol is used for collecting and updating
information about which resources participate in the
computation at any given time. The impossibility of guaranteeing
consistent views of group membership in asynchronous, unreliable
systems was proven in \cite{chandra}. Even in 
reliable systems, membership protocols are expensive, requiring
several phases for consistency. 

A group is defined as a set of members. It is initialized when the
first member enters the group and ceases to exist when the last member
leaves. A process joins a group by finding one or more members of the 
group and leaves it either by leaving or by failing. We assume the
existence of a fault-tolerant method by which 
processes can find other processes, such as broadcasting (when
applicable), known addresses of gossip servers (described below), or a
location  service. For the moment, we assume that gossip servers
exist. 

Our membership protocol is inspired 
by the failure-detection mechanism based on epidemic communication
presented in \cite{vanRenesse}. Other membership protocols based on epidemic
communication \cite{golding} are more
elaborate and introduce constraints or costs that are not
justified in our case.

The membership protocol works as follows: when a new computer joins
the group of resources, it sends its address to some known gossip
servers. The gossip servers act as any other member of the group,
except that at least one of them is guaranteed to be active at
any given moment during the computation. This is
a loose fault-tolerance constraint, easily achievable, without extra
costs, by increasing the number of gossip servers in the system. The
main task of these servers is to propagate information about the newly
arrived members.

Each member process maintains a view of group membership. The view
defines a set of processes that the member believes are part of the group
at any given time. In addition, it contains specific information
designed to log the members' activity by keeping track of 
when it last heard of each (known) member, directly from it or
through the 
gossip system. The parameters involved in this mechanism (for example,
the frequency of gossiping and the timeout period used to deduce
failure of a passive member) are chosen to keep
communication and the probability of false membership information
under some threshold values \cite{vanRenesse}.

Among the advantages of using this membership protocol are (1)
scalability in network load with the size of the group, (2) tolerance
to a small percentage of message loss or failed members, and (3)
scalability in accuracy with the number of members.

\subsection{Fault-Tolerance Mechanism}
\label{sec:FTM}

For B\&B algorithms, the loss of a subproblem is unacceptable when the
accuracy of the solution is important.

Our proposed fault-tolerance mechanism does not attempt to detect
failures of computers and restore their 
data, but rather focuses on detecting missing results. Given that the
B\&B tree of problems is 
dynamic, how is it possible to know the set of existing problems, so
that, knowing the problems completed, one can infer the set of not-completed
problems?

Our solution exploits the fact that the subproblems dynamically
generated by the 
B\&B algorithm are nodes of a tree. Each node can be uniquely
represented by its position in the tree. If we encode the position of
the nodes in the tree, we obtain a unique code for each
subproblem. Furthermore, given a set of nodes of the tree, we can
easily find its complement, that is, the list of nodes of the tree that
are not in the given set.

\subsubsection{Problem Representation}

Without loss of generality, we assume that the branching factor for
the search tree is 2 and that each branch is a decision on a condition
variable. Therefore, a subproblem is entirely described by a sequence
of pairs $\langle x_i, value \rangle $ where $x_i$ is a condition
variable and $value$ is $0$ or $1$, indicating the left or the right
branch, respectively. We need to include condition variables in the 
subproblem encoding because the order in which condition variables are
considered may vary over the tree.  For example, the left subtree of
a node that branches upon $x_k$ may consider $x_i$ first and
therefore will generate the subproblems $(\langle x_k, 1 \rangle,
\langle x_i, 0 \rangle) $ and $(\langle x_k, 1 \rangle, \langle x_i, 1
\rangle) $, whereas the right subtree may branch upon $x_j$ first,
producing the subproblems $(\langle x_k, 0 \rangle, \langle x_j, 0 
\rangle) $ and $(\langle x_k, 0 \rangle, \langle x_j, 1 \rangle) $.

\begin{figure}[ht]
\begin{center}
\includegraphics[angle=270,width=3in]{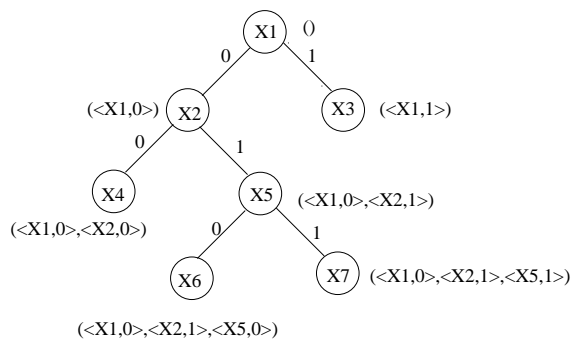}
\caption{\small Problem representation}
\end{center}
\end{figure}

Each pair $\langle x_i, value \rangle $ introduces and assigns a
condition to a new variable. That is what makes the codes
(subproblems) self-contained: the code (along with the initial data,
which is provided by a gossip server when a process joins the
computation) is enough to initiate a problem on any processor. 

\subsubsection{Mechanism Description}

Our failure-recovery mechanism allows each process to 
detect missing problems independently, based on local information
about completed problems.

\begin{figure}[ht]
\begin{center}
\includegraphics[angle=270,width=1.4in]{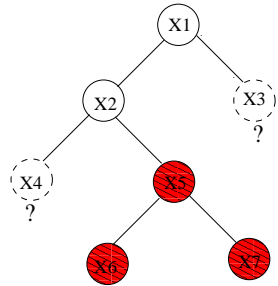}
\caption{\small Completed, unsolved, and solved problems: shadowed nodes represent
completed problems, dashed nodes represent unsolved problems (i.e.,
problems that are still in the active pool), and plain nodes represent
solved but uncompleted problems.}
\label{fig:completed}
\end{center}
\end{figure}

We consider a subproblem \textit{solved} after the branching operation
has been performed on it. \textit{Solved} subproblems are not
necessarily \textit{completed}: 
we consider a subproblem to be \textit{completed} if it is solved
and either it is a leaf or both its children are completed (see
Figure \ref{fig:completed}). 

Every process maintains a list of new locally completed subproblems and a
 table of the completed problems it knows about. When a problem is
 completed, it 
is included in the local list. When $c$ problems (codes) are in the
list or the list has not been updated for a long time, it is sent to
 $m$ of the other members as a \textit{work report}  
message. When a member receives a work report, it stores the report in its
 table. Occasionally, in order to inform new members of the current state
of the execution and to increase the degree of consistency, a member
 sends its table of completed problems to a randomly chosen member. 

The size and the number of the problem codes vary with the shape and
number of nodes of the B\&B tree. The deeper the node in the tree, the
larger the size of its code; the more nodes in the
tree, the larger the number of codes. Since the completion of a parent
node implies the completion of its children, communication costs can
be reduced by compressing work 
report messages, via the recursive replacement of pairs of
sibling codes with the code of their parent, and the deletion of codes
whose ancestors are also in the list. Simulations performed on real
B\&B trees confirmed that the compression rate is better when
processors are sufficiently loaded: the taller the subtree 
completed locally, the larger the number of codes that do not need
to be sent. 

Failure recovery is achieved as follows. When a member runs out of
work and an attempt to get work through 
the load-balancing mechanism fails, it chooses an uncompleted problem
(by complementing the code of a solved problem whose sibling is not solved)
and solves it. The mechanism ``repairs'' system failures due to, for
example, a computer that failed before sending work reports or work
reports that were lost before reaching any machine. Note that this
mechanism also works in the case of temporary network partitions.

This simple, fully distributed mechanism can lead to redundant
work in two situations: (a) the lag in updating information can lead
to faulty presumptions on failure; and (b) the lack of coordination among
processors permits multiple members to work on 
the same problem. The former case can be fixed easily by interrupting
the redundant work when information is updated. The costs of the
latter situation can be reduced by employing more sophisticated
methods for choosing work, such as using the location of the last problem
completed locally. Notice, however, that redundant computation may be
inevitable. 

If information about completed problems is spread uniformly, then the loss
of a percentage of members may not lead to information loss: if the
information about the problems reported to be completed still exists
in the system, they will not have to be redone.

\subsection{Almost Implicit Termination Detection}
\label{sec:TD}

The problem encoding used for implementing the fault-tolerance
mechanism also has the advantage of implicitly solving the termination
detection problem. When successive code compressions of local lists and
tables lead to the code of the root problem, termination is
detected. Since none of the communication mechanisms used guarantees
data consistency, it is possible that some members do not have enough
information to detect termination. That is why, before
termination, each member that detected the termination will have to
send one more work report, that is, the code of the root problem, to all
members from its local membership list. 

\subsection{Comparison with DIB}
Both DIB and the algorithm we propose are decentralized and
fault-tolerant algorithms that work on a dynamic, tree-like search
space. Both 
algorithms implement low-cost, simple fault-tolerance protocols for
the price of potentially redundant work. However, the two algorithms
have different failure-recovery mechanisms and  react differently in
the case of failure.

DIB uses a hierarchical structure for
failure detection and recovery that imposes the need for a reliable or
duplicated node for the root of this hierarchy. Moreover, the failure
of a node affects not only the problems solved locally and not
reported as solved yet, but also the problems given to other nodes,
whose completion cannot be reported (and therefore considered)
anymore. 

In our algorithm, all processes are equally responsible for the
behavior of the system in case of failure. Our simulation studies
confirm that the failure of all processes but one still allows the
problem to be correctly solved. The mechanism is also reliable in the
case of faulty network links or temporary network partitions. 

However, the homogeneity involved in our algorithm has a communication
cost: information about the completion of a problem is eventually
spread to all processes, directly (by reporting the code of the
problem) or indirectly (by reporting the completion of one of its
ancestors).

Performance comparisons of DIB and our
algorithm are of limited interest for two reasons. Because DIB was
designed for a 
wide range of applications, such as recursive backtrack, alpha-beta
search and branch-and-bound, its speedup is ``excellent for exhaustive
traversal and quite good for branch-and-bound''
\cite{dib}. Furthermore, speedup 
results are given for maximum 16 processors, while we are interested
in many more resources. 

\section{Experimental Studies}

\label{sec:Simulation}
We use simulations rather than a real implementation to evaluate our
algorithm, as the use of simulation techniques provides great
flexibility in testing a wide range of B\&B strategies in a 
variety of Internet-like environments.

\subsection{Experimental Goals}

The goals of our experimental work are as follows: (1) to verify
reliability and evaluate 
the overall performance of the algorithm, focusing on the costs
introduced by the fault-tolerance mechanism; and (2) to evaluate
scalability for different problem  classes and environments. Our work
to date has focused primarily on the first of these two issues.

We studied algorithm reliability by testing various failure
scenarios. The costs introduced by our fault-tolerance
mechanism are communication costs, storage space, tree contraction
time, and redundant work. Because we avoid centralized control by
spreading information throughout the system, communication costs may
be significant. Redundant work may increase when communication
conditions are poor (messages are delayed or lost) or when work load is
low. Storage space
may become a serious concern for large problems because the algorithm
permits (and benefits from) the replication of data. However, the
results we obtained encourage us to continue our research in this
direction.

\subsection{Simulation Framework}

We used Parsec \cite{parsec} to develop our simulation system. Parsec
is a C-based simulation language for 
sequential and parallel execution of discrete-event simulation
models. Processes are modeled by objects; interactions among objects
are modeled by  time stamped message exchanges. 

Our simulation system incorporates a detailed representation of load
balancing, failure recovery, and termination detection mechanisms. We
do not include yet the membership protocol: hence, the pool 
of resources is predetermined and varies only with failures. Each
process, after it has solved a B\&B subproblem, checks to see whether
any messages are pending. If it received a work request, it
satisfies the request if there are enough problems in its active
pool. If it received a work report, it merges that report with its local
information on completed problems and contracts the result.  

The simulation was configured so that it could be driven
either by \textit{real} (precomputed) B\&B trees or by random trees. For
real problems, we tested our algorithm on a set of \textit{basic
trees} that we obtained from an instrumented B\&B code. Basic trees are trees generated by executing a
branch-and-bound  algorithm without eliminating the unpromising
nodes.

For each node in the tree, we have the following information: 
(1) the node identifier, (2) its bound value, (3) the time needed for
computing the bound value and expanding the node or determining
infeasibility, and (4) a value specifying whether the bound
value is a feasible solution. The bound values are used for
pruning the test tree and obtaining the B\&B tree, and for
computing the optimal solution. The time value is used for simulating
the execution time needed for the bounding operation. Notice that the time
values determine the granularity of the subproblems. During our
experiments, we tuned this granularity by multiplying all time values
by a constant factor, and we studied how granularity affects the
overall performance of the B\&B algorithm. 

Running simulations on basic trees leaves enough room for generating
different B\&B trees, depending on communication
characteristics (for example, up-to-date information about the best-known
solution influences pruning decision) and on the number of
processors (because the number of nodes expanded may vary with the
number of processors). Note that the basic  branch-and-bound operation
\textit{decompose} is recorded within the basic tree structure.

Because the amount of communication and storage space depends on the
shape and the size of the tree, testing trees resulted from solving real
problems provides better accuracy. However, creation of basic
trees is computationally infeasible for anything but small
problems. But for testing reliability, and later scalability, the
number of nodes is the only important feature of  the test
tree. Therefore, we enriched our set of test trees with randomly
created trees of various sizes and tested them without eliminating the
unpromising nodes.

\subsection{Results}
\label{sec:Results}

Our simulator measured execution time, communication
costs, and storage space. We tested the algorithm on relatively small problems
(up to tens of thousands of nodes expanded), with no optimization efforts: work
reports are sent to randomly chosen resources, without eliminating
redundant messages. When out of work, resources ask 
randomly chosen resources for work, without using previous experience
to increase performance.

\subsubsection{Algorithm Performance}

\begin{figure}[htp]
\begin{center}
\includegraphics[width=5in]{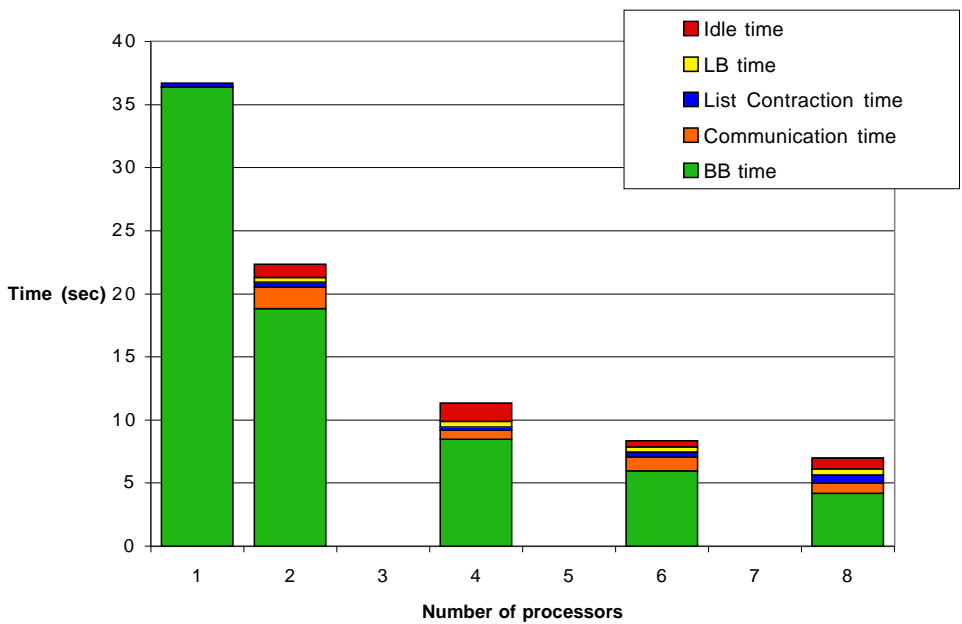}
\caption{\small Execution time for solving a real problem with
approximately 3,500 expanded nodes. In this problem, average node cost
is 0.01 sec; communication costs are modeled as $ 1.5+0.005 \times
L $ ms. for messages of size $L$ bytes.} 
\label{fig:hist_p0201}
\end{center}
\end{figure}

Figure \ref{fig:hist_p0201} shows 
results obtained for a small problem (approximately 3500 nodes
expanded) with average granularity of 0.01 seconds per node. For this 
problem, the overhead introduced by the algorithm reaches 36\% for 8
processors. This is determined by three factors: (1) the
relatively high communication costs considered ($1.5+0.005 \times
L$ milliseconds for messages of size $L$ bytes); (2) the cost of
the dynamic load balancing mechanism for a network of workstations
\cite{kumar-scalability}; and (3) the small 
granularity of the subproblems. We will see that for a larger problem
(Table \ref{tlb:trim4}) the overhead is much lower (15.58\% 
of the total execution for 100 processors, from which 13.67\% are load
balancing costs, 0.78\% communication time and 1.13\% list contraction
time).  Furthermore, this 
overhead can be controlled by tuning various execution parameters. For
example, less frequent termination verification leads to lower list
contraction costs but may increase idle time. Sending work reports
more rarely may decrease communication time and list contraction costs
but may increase termination detection time, because of lack of
information. If the failure recovery mechanism is activated  (decides
that a problem was lost and recreates it) less often, the overhead
introduced (list contraction and redundant work costs) is lower, but
recovery in case of failure is also slower. 

The tests we performed on larger problems (total
uniprocessor execution time of around 75 hours) show that communication and
storage space costs remain negligible (Table \ref{tlb:trim4}). We find
that good performance is achieved on up to 100 processors. These
preliminary results encourage us to continue evaluating our algorithm
on larger problems, with larger number of resources. 

  \begin{table}[htp]  
  \centering
  \bigskip
    \begin{tabular}{|r|r|r|r|r|r|r|}
    \hline
    No.  & Execution & B\&B  & Contraction & \multicolumn{2}{|c|}{Storage Space} & Communication \\
Processors & Time  & Time & Time  & Total &Redundant  & \\
 & (hours) & (\%) & (\%) & (MB) & (MB) &(MB/hour/processor)\\
    \hline
     10 & 7.93 & 98.11\% & 0.35\% & 0.42 &0.16 & 1.01\\
     30 & 2.91 & 90.42\% & 5.20\% & 3.76 &1.92 & 1.40 \\
     50 & 2.00 & 81.19\% & 11.73\% & 12.65 &6.43 & 2.34\\
     70 & 1.37 & 87.32\% & 2.33\%& 19.81 &10.13 & 3.16\\
     100 & 1.04 & 84.40\% & 1.13\% & 43.06 &21.88 & 4.56\\
    \hline

    \end{tabular}
  \caption{\small Simulated execution of a real problem (approximately 79,600 nodes
  expanded). In this problem, average cost per node is 3.47
  sec. Communication costs are modeled as $ 1.5+0.005 \times 
L $ ms. for messages of size $L$ bytes.}
\label{tlb:trim4}
  \end{table}

\begin{figure}[!htp]
\begin{center}
\includegraphics[angle=270,width=3.2in]{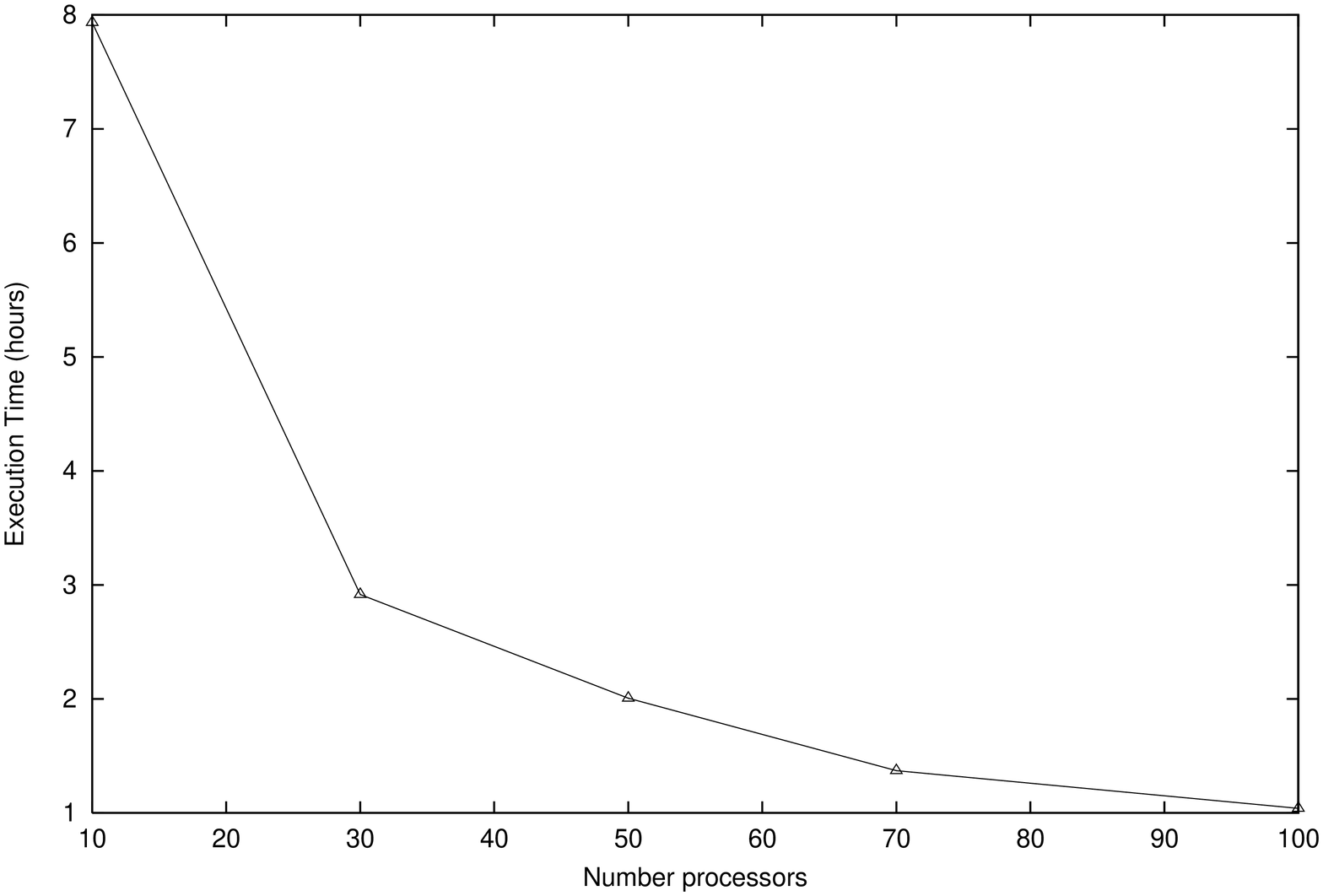}
\includegraphics[angle=270,width=3.2in]{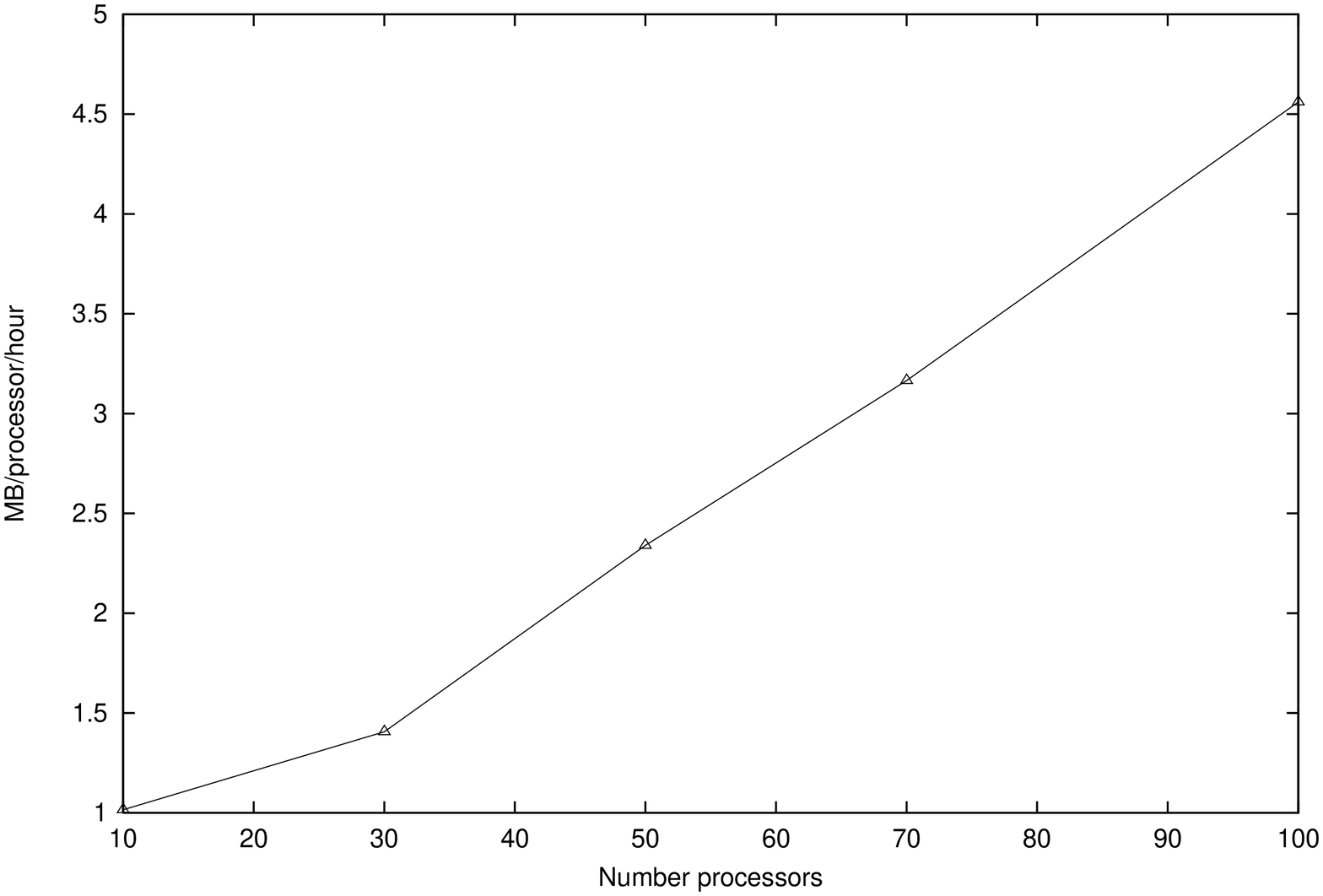}
\caption{\small Speedup and communication for the execution of the
problem from Table \ref{tlb:trim4}.}
\end{center}
\end{figure}

Communication per processor increases with the number of processors
because the number of work reports sent per processor increases:
since the work load is lower, and therefore processes are idle longer
periods of time, they suspect termination and send more work
reports. Storage space is measured for the entire system. The results
obtained---43 MB storage space for 100 machines---are promising. 

A normal trend would be that the amount of time spent on
list contraction increases with the number of processors, since the
number of messages circulated within the system increases and the
receiving of a work report message requires a list contraction
procedure. But because this depends on how the subproblems are
assigned on processors, a lucky configuration may lead to unexpected
good results (as for 100 processors, Table \ref{tlb:trim4}).

The amount of redundant work performed is another interesting
measure of our algorithm that remains to be evaluated. However, this amount
can be reduced by tuning parameters (for example, how soon failure is 
suspected after a machine unsuccessfully tries to get work) or by
designing more sophisticated methods for picking up unsolved
problems. 

When varying problem granularity (by multiplying the time needed
to solve a problem with some constant values), we observed the
following (not unexpected) behavior: The number of nodes expanded may
vary, because the information of the best-known solution is computed
at different moments. Load balance is better when
granularity is coarser. Communication increases unnecessarily because
work reports are sent at fixed time intervals. This last observation
taught us that for scalability, we need to design an adaptive
mechanism for deciding how often work reports should be sent, based on
information collected at runtime: for example, information about
execution time per subproblem and frequency of messages received. 

\subsubsection{Fault Tolerance}

Because our termination detection mechanism operates by detecting that
all expanded problems have been completed, it is straightforward to
verify that our fault-tolerance algorithm is working correctly---we
simply verify that termination is detected. For visualizing the
behavior of the 
algorithm, we used Jumpshot, a graphical visualization tool for \texttt{clog}
log file format. We used the MPE library developed by the MPICH team at
Argonne National Laboratory for logging the execution profile. 

\begin{figure}[!htp]
\begin{center}
\includegraphics[angle=270,width=6in]{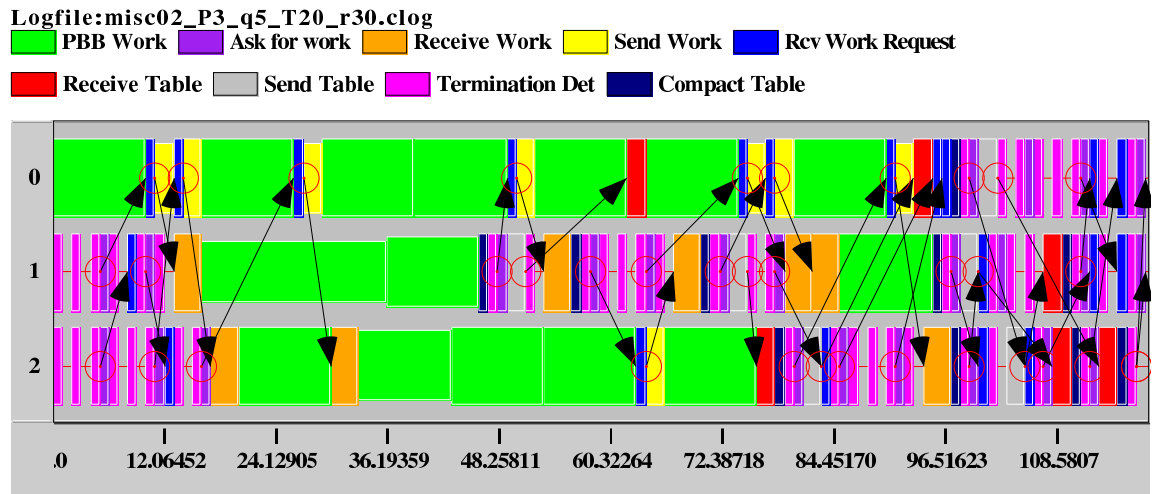}
\caption{\small Solving a very small problem.}
\label{fig:misc02_nf}
\end{center}
\end{figure}

\begin{figure}[!htp]
\begin{center}
\includegraphics[angle=270,width=6in]{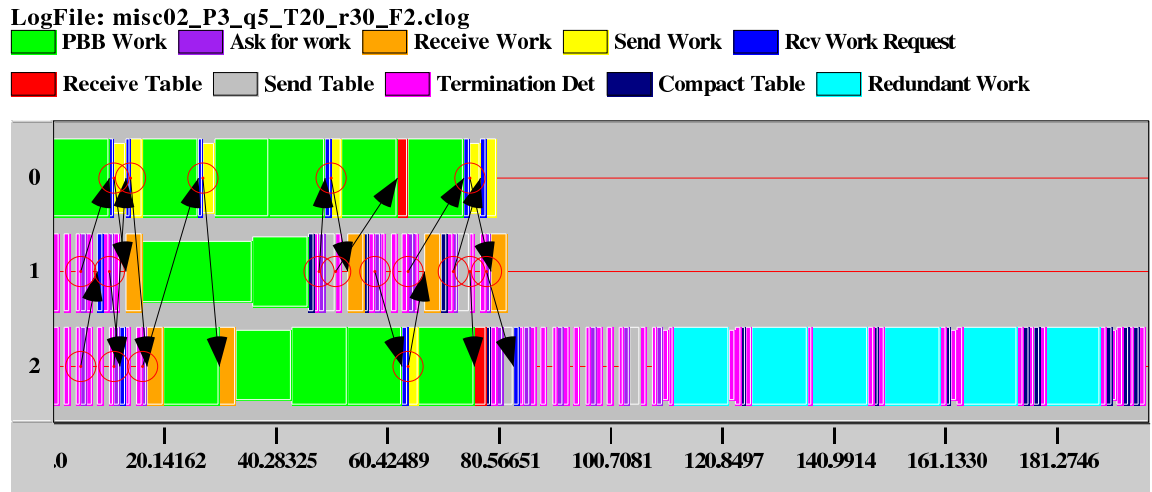}
\caption{\small The same problem: two processors crash about the
same time; the third processor recovers the lost work.}
\label{fig:misc02_wf}
\end{center}
\end{figure}

Figures \ref{fig:misc02_nf} and \ref{fig:misc02_wf} are snapshots of
the execution of the algorithm on a very small problem. Figure
\ref{fig:misc02_nf} shows the behavior of the algorithm in the absence
of failures. The same problem is presented in Figure
\ref{fig:misc02_wf}, where two of the three processors fail at about
85\% of the execution time. The only processor available after this
moment is able to solve the problem and terminate.  

\section{Conclusions and Future Work}
\label{sec:Conclusions}

We presented a failure-recovery mechanism suited for a tree-like problem
space. This mechanism and a low-cost group membership
protocol are the ingredients that transform a rather conventional 
parallel branch-and-bound algorithm into a scalable,
reliable, more powerful algorithm, able to exploit the
computational power of hundreds of Internet-connected
resources. Scalability is achieved through a fully distributed
design. The algorithm is fault tolerant under our assumptions
and can execute and terminate correctly even if only a single resource
remains available.  

We solved the difficult problems of fault tolerance and termination
detection in distributed environments by exploiting problem-specific
features, specifically the tree structure of the problem
space. While the mechanism we propose is not applicable to
all distributed computations, we believe that a large class of
problems can benefit from it. 

We have used simulation studies to explore the behavior of our
algorithm. Initial results on relatively small problems and up to 100 
processors are promising: performance is good despite the lack 
of optimization. Communication costs are reasonable, storage space costs are
negligible. However, we need results on a much larger number of
processors. We plan to introduce the group membership
protocol into our simulations and to test the algorithm under various
network conditions. An interesting issue to study is how the network 
characteristics influence the performance of the algorithm in general
and the costs introduced by the failure-recovery mechanism in
particular. Also, in order to accurately analyze scalability issues,
we plan to design a flexible scheme for adapting parameters to runtime
informations, such as total execution time and execution time per problem.


\begin{thebibliography}{99}

\bibitem{alon} N. Alon, A. Barak, U. Manber. 1987. On disseminating
information reliably without broadcasting. \textit{Proceedings of the
7th International Conference on Distributed Computing Systems} IEEE
Computer Society Press, 74--81.

\bibitem{isis} K. Birman. 1986. ISIS: A System for Fault-Tolerant
Distributed Computing. Technical Report TR86-744. Cornell University,
Computer Science Department.

\bibitem{chandra} T. Chandra, V. Hadzilacos, S. Toueg,
B. Charron-Bost. 1996. On the impossibility of group
membership. \textit{PODC '96. Proceedings of the 15th Annual ACM
Symposium on Principles of Distributed Computing}, 322--330.

\bibitem{demers88} A. Demers, D. Greene, C. Hauser, W. Irish,
J. Larson, S. Shenker, H. Sturgis, S. Swinehart,
D. Terry. 1988. Epidemic algorithms for replicated database
maintenance. \textit{Operating Systems Review}, 22(1):8--31.

\bibitem{eckstein94} J. Eckstein. 1994. Parallel branch-and-bound algorithms
for general mixed integer programming on the CM-5. \textit{SIAM
J. Optimization}, 4(4):794--814. 

\bibitem{eckstein-comm} J. Eckstein. 1997. How much communication
does parallel branch and bound need?. \textit{INFORMS Journal on
Computing}, 9(1):15--29.

\bibitem{farley80} A. Farley. 1980. Minimum-time broadcast networks,
\textit{Networks}, 10:59--70.

\bibitem{dib} R. Finkel, U. Manber. 1987. DIB - A distributed
implementation of backtracking. \textit{ACM Transactions on Programming
Languages and Systems}, 9(2):235--256.

\bibitem{fischer} M. Fischer, N. Lynch, M. Paterson. 1985. Impossibility of
distributed consensus with one faulty process. \textit{Journal of the
ACM}, 32(2):374--382.

\bibitem{TheGrid} I. Foster, C. Kesselman. 1999. \textit{The Grid ---
Blueprint for a New Computing Infrastructure}. Morgan Kaufmann
Publishers. 

\bibitem{crainic94} B. Gendron, T. Crainic. 1994. Parallel
branch-and-bound algorithms: Survey and synthesis. \textit{Operations
Research}, 42(6):1042--1066.

\bibitem{cristian} F. Cristian. 1991. Understanding fault-tolerant distributed
systems. \textit{Communications of the ACM}, 34(2):56--78.

\bibitem{cris-fetzer} F. Cristian, C. Fetzer. 1999. The timed asynchronous
distributed system model. \textit{IEEE Transactions on Parallel and
Distributed Systems}, 10(6):642--657.

\bibitem{gartner} F. Gartner. 1999. Fundamentals of fault-tolerant
distributed computing in synchronous environments. \textit{ACM
Computing Surveys}, 31(1):1--26.

\bibitem{golding} R. Golding, K. Taylor. 1992. Group membership in
the epidemic style. Technical Report UCSC-CRL-92-13, University of
California at Santa Clara.

\bibitem{hahn} P. Hahn, W. Hightower, T. Johnson,
M. Guignard-Spielberg, C. Roucairol.  1999. Tree elaboration strategies in
branch-and-bound algorithms for assigning the quadratic assignment
problem, \textit{6th SIAM Conference on Optimization}. %PAGES

\bibitem{lenstra} A. Lenstra. 1995. Factoring integers using the Web and the
number field sieve. Tech. report, Bellcore.

\bibitem{kumar-scalability} V. Kumar, A. Grama, V. Rao. 1994. Scalable load
balancing techniques for parallel computers. \textit{Journal of Parallel
and Distributed Computing}, 22(1):60--79.

\bibitem{laforest97} C. Laforest. 1997. Broadcast and gossip in
line-communication mode. \textit{Discrete Applied
Mathematics}, 80:161--176. 

\bibitem{condor} M. Litzkow, M. Livny, M. W. Mutka. 1988. Condor---a
hunter of idle workstations. In \textit{Proc. 8th Int'l Conf. of
Distributed Computing Systems}, 104--111.

\bibitem{condor-ckpt} M. Litzkow, T. Tannenbaum, J. Basney,
M. Livny. 1997. Checkpoint and migration of UNIX processes in the Condor
distributed processing system. Technical Report \#1346, University of
Wisconsin-Madison, Computer Sciences Department. 

\bibitem{legion}  A. Nguyen-Tuong, A. Grimshaw. 1998. Using Reflection
for Incorporating Fault-Tolerance Techniques into Distributed
Applications. Tech. report CS-98-34, University of Virginia.  

\bibitem{parsec}  UCLA Parallel Computing Laboratory. Parsec ---
Parallel Simulation Environment for Complex
Systems. \texttt{http://may.cs.ucla.edu/projects/parsec}.

\bibitem{chkpt-mngt} J. Pruyne, M. Livny. 1996. Managing checkpoints
for parallel programs. Workshop on Job Scheduling Strategies for
Parallel Processing IPPS '96. 

\bibitem{vanRenesse} R. van Renesse, Y. Minsky, M. Hayden. 1998. A
gossip-style failure detection service, IFIP International Conference
on Distributed Systems Platforms and Open Distributed Processing
(Middleware'98).

\bibitem{schneider} F. Schneider. 1993. What good are models and what models
are good? \textit{Distributed Systems (2nd ed.)}, S. Mullender,
Ed. Addison-Wesley Longman Publ. Co., Inc., Reading, MA, 17--26.

\bibitem{ft-globus} P. Stelling, I. Foster, C. Kesselman, C. Lee,
G. von Laszewski. 1998. A fault detection service for wide area distributed
computations. \textit{Proc. 7th IEEE Symp. on High Performance
Distributed Computing}, 268--278.

\bibitem{hjk} V. Zandy, B. Miller, M. Livny. 1999. Process
hijacking. \textit{Proc. 8th IEEE Symp. on High Performance 
Distributed Computing}, 177--184.





\bibitem{BBtaxonomy} H. Trienekens, A. de Bruin. 1992. Towards a taxonomy of
parallel branch-and-bound. Report EUR-CS-92-01, Erasmus University
Rotterdam.

\end{thebibliography}
\end{document}